\documentclass[12pt]{article}
\pdfoutput=1

\usepackage{amsmath,amssymb,amscd}
\usepackage{listings}
\usepackage{caption}
\usepackage{dsfont}
\usepackage{slashed}
\usepackage{color}
\usepackage{ulem}
\usepackage[pdftex]{graphicx}
\usepackage{epstopdf}
\usepackage{subfigure}
\usepackage{epsfig}
\usepackage{listings}
\usepackage{caption}
\usepackage{cite}
\usepackage{multirow}

\newcommand{\beq}{\begin{equation}}
\newcommand{\eeq}{\end{equation}}
\newcommand{\nbea}{\begin{align*}}
\newcommand{\neea}{\end{align*}}
\newcommand{\nbeq}{\begin{equation*}}
\newcommand{\neeq}{\end{equation*}}

\usepackage{multirow}
\usepackage{array}
\newcolumntype{M}[1]{>{\centering\arraybackslash}m{#1}}
\newcolumntype{N}{@{}m{0pt}@{}}

\begin{document}

\pagestyle{empty}

\baselineskip=21pt

\begin{center}

{\large {\bf Size of Shell Universe in Light of {\it Fermi} GBM Transient Associated with GW150914}}

\vskip 0.2in

{\bf Merab Gogberashvili}\textsuperscript{a,b},~%\footnote{luca.bonetti696@gmail.com},
{\bf Alexander S. Sakharov}\textsuperscript{c,d,e},~\\%\footnote{alexandre.sakharov@cern.ch},
{\bf Edward K. Sarkisyan-Grinbaum}\textsuperscript{e,f},~\\%\footnote{sedward@mail.cern.ch},\\

\vskip 0.2in

{\small {\it

\textsuperscript{a}{\mbox Department of Exact and Natural Sciences, Javakhishvili Tbilisi State University}\\
\mbox{Tbilisi 0179, Georgia}\\
\vspace{0.25cm}
\textsuperscript{b}Department of High Energy Physics, Andronikashvili Institute of Physics\\
\mbox{Tbilisi 0177, Georgia}\\
\vspace{0.25cm}
\textsuperscript{c}Department of Physics, New York University\\
4 Washington Place, New York, NY 10003, United States of America\\
\vspace{0.25cm}
\textsuperscript{d}Physics Department, Manhattan College\\
{\mbox 4513 Manhattan College Parkway, Riverdale, NY 10471, United States of America}\\
\vspace{0.25cm}
\textsuperscript{e}Experimental Physics Department, CERN, CH-1211 Gen\`eve 23, Switzerland \\
\vspace{0.25cm}
\textsuperscript{f}Department of Physics, The University of Texas at Arlington\\
{\mbox 502 Yates Street, Box 19059, Arlington, TX 76019, United States of America} \\}}

\vskip 0.2in

{\bf Abstract}

\end{center}

\baselineskip=18pt \noindent

%%%%%%%%%%%%%%%%%%%%%%%%%%%%%%%%%%%%%%%%%%%%%%%%%

%\begin{abstract}
The possible gamma ray burst occurred in location and temporal consistence 
with gravitational wave event GW150914, as reported by {\it Fermi} Gamma-ray Burst Monitor (GBM), offers 
a new way of constraining models with extra dimensions.
Using the time delay in arrival of the gamma ray transient observed by {\it Fermi} GBM 
relative to the gravitational waves event triggered by the LIGO detectors 
we investigate the size of the spherical brane-universe expanding in multi-dimensional space-time. 
It is shown that a joint observation of gravitational waves in association with gamma ray
burst can provide a very stringent bound on the spatial curvature of the brain.

\vskip 3mm
PACS numbers: 04.30.Nk; 98.70.Rz; 11.25.Uv
\vskip 1mm

Keywords: Gravitational waves; Gamma ray burst; Extra dimensions
%\end{abstract}
\vskip 5mm

\leftline{November 2016}

\vskip 10mm

%%%%%%%%%%%%%%%%%%%%%%%%%%%%%%%%%%%%%%%%%%%%%%%%%%%%%%%%%%%%%%%%%%%%%%%%%%%%

Recently, the team of the Laser Interferometer Gravitational-Wave Observatory (LIGO) has reported
on simultaneously detected identical chirp like signals in their both detectors \cite{ligoGW}.
This historical event, which became world wide known as GW150914, has been interpreted as the
first detection of gravitational waves, predicted by the general relativity,  from inspiral and
merger of a pair of black holes (BHs) at a luminosity distance of about $0.4$ Gpc. The inferred merger
of the BH binary whose components have  masses $M_1=36_{-4}^{+5}M_{\odot}$ and
$M_2=29_{-4}^{+4}M_{\odot}$ has formed the final BH of mass $M=62_{-4}^{+4}M_{\odot}$ releasing
$E_{\rm gw}=3.0_{-0.5}^{+0.5}M_{\odot}/c^2\simeq 5.4\cdot 10^{54}$~erg in gravitational waves.  

The {\it Fermi} Gamma-ray Burst Monitor (GBM), being exposed consistently with the direction of GW150914, 
revealed, at about $3\sigma$-level, the presence of a weak hard X-ray transient signal of luminosity 
$E_{\gamma}\simeq 10^{49}$~erg at photon energies 10 and 1000~keV over 1~s that appeared $0.4$ seconds 
after the gravitational wave event \cite{fermiTransient}. Although having ill-constrained localization, 
because of non optimal pointing of the GBM detector, it was suggested to associate this short lasted 
about $1$ second transient with GW150914. The properties of this {\it Fermi} GBM transient are broadly 
consistent with a short gamma ray burst (SGRB), apart the fact that it would be significantly harder 
than the typical $E_{\rm peak}-E_{\rm iso}$ relation for SGRBs \cite{china1LigoFermi}. We notice that 
so far no other instrument could report on detection of an electromagnetic transient counterpart or an 
afterglow \cite{int, des, fermiLAT, swift, agile, xmm, rew1}. Also, no success has been achieved in high 
energy neutrino follow-up of GW150914 \cite{icecube,rew1}. 
In addition, later on, two other gravitational waves event  
candidates from compact object
mergers have been reported by LIGO, namely GW151226 
and the subthreshold LIGO-Virgo Trigger LVT151012. 
Eventually, GW151226 has been confirmed to occur due to the inspiral and merger binary black
holes system, consisting of $14.2_{-3.7}^{+8.3}M_{\odot}$ and
$7.5_{-2.3}^{+2.3}M_{\odot}$ components at the luminosity distance 
of $0.4$~Gpc~\cite{GW151226}. The subthreshold trigger has not been 
confirmed as being associated 
with a real astrophysical event. No electromagnetic conterpart of GW151226 
has been revealed either by gamma ray observations performed by 
{\it Fermi} GBM and LAT~\cite{Fermi_GW151226} or optical follow up 
campaigns~\cite{Opticsi_1_GW151226,Opticsi_2_GW151226}.
With many more LIGO events expected in the 
future, it would be straightforward to test whether GRBs are a common byproduct of BHs mergers.

Right after the discovery of the gravitational wave signal, some tests on general relativity have been
performed \cite{ligoGR, sibiryacovVelocity}, including measurements of the velocity and mass of the graviton
as well as violations of Shapiro delay and Einstein's equivalence principle between the gravitons at different
frequencies \cite{shapiro}. It has been also understood that the putative association of the {\it Fermi} GBM transient 
with GW150914 can be compiled in spectacularly useful results on general relativity, quantum gravity and astrophysics.
In particular, time lag between arrival of GW150914 event and the {\it Fermi} GBM transient has been used
in \cite{ellisLigoFermi, china1LigoFermi} to obtain the most stringent constraints on the velocity of the graviton.

Before detection of GW150914 putatively followed by the X-ray transient there was no specific astrophysical 
analysis on the electromagnetic radiation counterparts from the merger of binary stellar BHs. 
However, provided that the BHs have spin, as seems inevitable, and there are relic magnetic fields and disk debris remaining from the formation of the BHs or from their accretion history, one can substantiate that merging BHs potentially provide an 
environment for gamma rays emission~\cite{Silk,jap} and even for accelerating cosmic rays to ultrahigh energies 
\cite{Silk}. Indeed, simulations of gas and magnetic fields around the merging systems suggest that the motion
of two BHs in a magnetically dominated plasma could generate a magnetosphere and nebular structure similar to those 
inferred in pulsars, as well as collimated jets~\cite{plasma1, afterglow-1, afterglow-2, afterglow-3, 
afterglow-4, afterglow-5, afterglow-6, afterglow-7}. 

The power of the gamma emission generated is however quite uncertain, and 
in general depends on parameters and unknown structural details of the system.
Most models \cite{afterglow-1, afterglow-2, afterglow-3, afterglow-4, afterglow-5, afterglow-6, afterglow-7} are 
in line with the so called Blandford-Znajek (BZ) process~\cite{firsJets} that extracts the space-time 
rotational energy of the BHs to generate a powerful electromagnetic outflow in externally 
supported magnetic field. On the cost of rescaling the BH 
mass and the magnetic field, it seems that the same mechanism can be applied to stellar BHs. Poynting
flux for such kind of emission has been estimated in \cite{lyut1} and reads
\begin{equation} \label{L_bz}
L_{\rm BZ} = \frac{(GM)^3B^2}{c^5R}\simeq 3.2\cdot 10^{46}{\rm erg\, s^{-1}}\left(\frac{M}{100M_{\odot}}\right)^3
\left(\frac{B}{10^{11}{\rm G}}\right)^2\left(\frac{R_S}{R}\right) ~,
\end{equation}
where $M$ is the final BH mass and $B$ is the strength of the external magnetic field. The orbital radius $R$ can 
be taken to be equal to the Schwarzchild radius $R_S = 2GM/c^2$ $\simeq 3\cdot 10^5 (M/M_{\odot})$~cm. 

Thus, if an external magnetic field of the order of $\gtrsim 5\cdot 10^{12}$~G could be generated, equation (\ref{L_bz}) 
implies  that the BZ process would extract enough electromagnetic 
luminosity to  account for the level of emission of the transient reported by {\it Fermi} GMB. Although a magnetic field 
of such strength can be easily accounted for pulsars and magnetars, there is a certain concern on an ability of 
ambient plasma surrounding a BH to anchor the magnetic field of such strength \cite{lyut2}. Mostly, the concern is 
related to the fact that plasma  in a close vicinity of the horizon of a resulting BH is expected to be diluted, 
since a BHs binary hollows out from material the inner few hundreds of Schwarzschild radius before the merger 
\cite{plasma1}. However, as pointed out in \cite{masaros}, a small disk or celestial body may be involved in the 
binary of two BHs. In this case, magnetic field of required level would be supplied by the mini-disk via 
magnetohydrodynamic instabilities such as the magneto-rotational instability \cite{masaros}. Also, as argued 
in \cite{vachaspati1} it would take infinite time, as measured by an external observer, for infalling matter to arrive 
at the horizon of a BH. Therefore, at the time of merger the accreted matter should be accumulated in the vicinity 
of the horizon forming so called "frozen" or "black" star \cite{blackStar1}. Thus, due to presence of significant 
amount of matter directly at the horizon which has been accreted with an angular momentum the magnetic field may be
compressed in the shell of the "black" star and amplified up to the level of neutron star or magnetar making the BZ 
mechanism working effectively \cite{blackStar2}. A generic version of BZ mechanism with respect to 
the association of the {\it Fermi} GBM transient and GW150914 is discussed in~\cite{genBZ}.   
Other models, outside the lore of the BZ process, are also discussed 
in the literature. In \cite{loeb}, it has been suggested that the merging BHs might have been generated in course 
of a collapse of a rapidly rotating massive star, that at the end a GRB occurs from a jet that originated in 
the accretion flow around the remnant BH. Another possibility of having in a binary system two gravitationally 
collapsing objects with non-vanishing electric charge has been discussed in \cite{reconnection}. 
In this case, the compenetration of the two magnetospheres occurring during the coalescence, 
through turbulent magnetic reconnection, produces a highly collimated
relativistic outflow that becomes optically thin and can power a SGRB. More exotic possibility of generation of
an electromagnetic counterpart of the GW150914 due to the appearance of a short living naked singularity during
the merger has been discussed in \cite{singularity}.  

Future gravitation wave observations with larger coverage
from {\it Fermi} GMB or other gamma ray bursts orbiters should 
settle if the binary BHs mergers indeed are accompanied by 
a gamma ray emission. However, it might happen that due to 
observer angle effects, the association of gravitational 
waves events from BHs mergers will only be confirmed once
a reasonably large sample of gravitational waves and gamma 
ray observations has collected. Indeed, as one can see from the above discussion,
GRB models usually invoke jets emitted along the
rotation axis of their progenitor BH. Thus, we can assume, due to Doppler boosting,
that our viewing angle is within the opening angle of the jet, otherwise the 
electromagnetic emission would be substantially suppressed bellow the
limit of detectability. Therefore, the most favorable situation for both
gravitational waves detection and collimated electromagnetic emission 
is in case we view the binary system, perpendicular to the rotation plane. 
It is clear, that the gravitational waves signal should not have
a strong dependence on the viewing angle. The difference in gravitational waves 
signal to noise ration between on face and on edge line of sight to the rotation plane 
of the binary system can be accounted by factor $\sqrt{8}\approx 2.8$~\cite{ligoGW}.
Due to a known degeneracy between the inclination angle of the line of sight and distance, in general
all the inclination angles are allowed by gravitational waves data~\cite{ligoGW}. 
The observed gamma rays however suggest that we see the system close to on face to the rotation plane
line of sight. The latter might not be the case in the observational configuration of GW151226 and 
thus caused negative results of its electromagnetic follow up 
campaigns~\cite{Fermi_GW151226,Opticsi_1_GW151226,Opticsi_2_GW151226}.

In this note, we exploit the relative timing of GW150914 and {\it Fermi} GBM events to bound the spatial
curvature in a class of models with large extra dimensions, where matter particles and radiation are
localized on a brane while the gravity can propagate in the bulk outside the brane.

The main virtue in warped extra dimensional models is their ability to solve many long standing problems in particle
physics, for instance, the  hierarchy problem \cite{merabRS, RS}.
A particular class of the brane scenario is the shall-universe model, where the universe is represented
as a 3-shell expanding in a higher dimensional hyper-universe \cite{merabShellUniverse, Bubble, Go-Ma}.
Interestingly, the hyper-sphere is a simple space possessing a positive curvature. Recent analysis of CBM data,
combined with other astronomical observations, suggests that the universe is nearly flat, but possibly with
small positive curvature, i.e. finite. Moreover, some observational data such as the isotropic runaway of
galaxies, the deficiency in the first modes of the angular power spectrum and an existence of the preferred
frame in the universe support this model. The shell-universe model also predicts a correct value of the
redshift parameter that corresponds to the transition from cosmic deceleration to acceleration without
introduction of dark energy on the brane \cite{Bubble} and provides a natural mechanism for the local increment
of the brane tension, leading to the modified Newton's law at galaxy scales, alternative to galactic dark matter \cite{Go-Ma}.

Since within the shell-universe scenario, the Standard Model particles are localized on the 3D spherical brane,
the propagation of a radiation probe (light) between two points should be affected by the curvature of the universe.
The graviton being able to propagate outside the brain, can make a shortcut between these points, what would look
like if in our 3D world the graviton was traveling faster than light. In the simplest case of a spherically symmetric
3D shell, the distance covered by a radiation probe propagated between two points on the sphere should be represented by an arc
distance $L$ between the source and the observer. In the same time, the bulk shortcut of the graviton would connect
these points simply via span of the arc. Thus, measuring an extent $d$ of the shortcut (span) one can calculate
the distance excess $\Delta x$ accumulated due to the curvature of the brain,
\begin{equation}\label{deltaX}
\Delta x = L - d \approx R\alpha -2R\sin\frac{\alpha}{2} \approx R\frac{\alpha^3}{24}~.
\end{equation}
Here,
\begin{equation} \label{R=a}
R \propto a(t)
\end{equation}
is the bulk radial coordinate  of the expanding brane, which serves as the dimensionful cosmological scale factor for
the brane (shell) observer, and
\begin{equation} \label{alpha}
\alpha = \frac dR
\end{equation}
is the central angle subtended by an arc on the 3-shell between the points of the source and the observer.
Therefore, as soon we  get in our disposal a gravitational wave signal along with an electromagnetic
counterpart associated with it, the size of the expanding shell (\ref{R=a}) and hence the curvature of
the observable universe for the brane observer can be measured.

In the case of the {\it Fermi} GBM transient arrived $0.4$ seconds later in coincidence with GW150914, one can
constrain the distance excess accumulated during the propagation of the radiation probe,
\begin{equation}
\Delta x \lesssim 10^5 ~{\rm km}~.
\end{equation}
Then using the estimation for the distance to the source of GWs given by LIGO \cite{ligoGW}
\begin{equation}
d \sim 10^{22} ~{\rm km} \gg \Delta x~,
\end{equation}
from (\ref{deltaX}) and (\ref{alpha}) we estimate the curvature radius of the shell-universe:
\begin{equation} \label{R}
R = \sqrt{\frac {d^3}{24\Delta x}} \gtrsim 10^{30}~km~.
\end{equation}
Note that this value is much larger than the radius of the observable universe (Hubble sphere),
\begin{equation}\label{RH}
RH \gtrsim 10^7~,
\end{equation}
where the Hubble constant is expressed as $H = \dot R/R$, according to relation (\ref{R=a}).

One can write down the expression for spatial curvature density of the universe,
\begin{equation}
\Omega_K = 1 - \Omega_M - \Omega_{\Lambda} = - \frac{1}{R^2H^2}~,
\end{equation}
where $\Omega_M$ represents the sum of the density fractions of baryons and the dark matter,
while $\Omega_{\Lambda}$ stands for the density fraction of the dark energy.
Using Eq. (\ref{RH}) one arrives to the constraint:
\begin{equation}\label{result}
|\Omega_K| \lesssim 10^{-14}~.
\end{equation}
The bound (\ref{result}) is much stringent than the one deduced by {\it Plank}, $|\Omega_K| < 0.005$ \cite {Plank}.
Of course, the constraint obtained here to be considered a definitive as much as the {\it Fermi} GBM transient can
be regarded a firm observation of a photon flash in coincidence with GW150914.
Moreover, in light of negative results of the electromagnetic follow up campaigns of 
another, more resent, LIGO's gravitational waves event GW151226 
performed by {\it Fermi}  GRB and LAT 
as well as by other instruments~\cite{Fermi_GW151226,Opticsi_1_GW151226,Opticsi_2_GW151226}, 
the bound (\ref{result}) to be treated as a prospective one requiring an accumulation 
of a much more sizeable sample of confirming observations.
Indeed, provided that gravitational waves signals from compact mergers are not strongly dependent 
on the orientation of the rotation plane of the binary while prompt gamma rays are essentially not expected 
if we are not inside the jet opening angle one may not expect that only few gravitational 
waves signals followed up by electromagnetic observations can confirm or falsify the association of the {\it Fermi} GBM
transient with GW150914. 

%%%%%%%%%%%%%%%%%%%%%%%%%%%%%%%%%%%%%%%%%%%%%%%%%%%%%%%%%%%%%%%%%%%%
\vskip 3mm
\noindent
{\bf Acknowledgments:} Merab Gogberashvili acknowledges the hospitality of CERN TH Department, where this work has been done.
The work of Alexander Sakharov is partially supported by the US National Science Foundation under Grants 
No. PHY-1505463 and No. PHY-1402964.

%%%%%%%%%%%%%%%%%%%%%%%%%%%%%%%%%%%%%%%%%%%%%%%%%%%%%%%%%%%%%%%%%%%%%%%%%%%%


\begin{thebibliography}{99}

\bibitem{ligoGW} LIGO Scientific and Virgo Collaborations,
                {\it ``Observation of Gravitational Waves from a Binary Black Hole Merger"},
                Phys. Rev. Lett. {\bf 116} (2016) 061102, arXiv: 1602.03837 [gr-qc].

\bibitem{fermiTransient} V. Connaughton {\it et al.},
                        {\it ``Fermi GBM Observations of LIGO Gravitational Wave event GW150914"},
                        arXiv: 1602.03920 [astro-ph.HE].

\bibitem{china1LigoFermi} X. Li, F.W. Zhang, Q. Yuan, Z.P. Jin, Y.Z. Fan, S.M. Liu and D.M. Wei,
                         {\it ``Implication of the association between GBM transient 150914 and LIGO Gravitational Wave event GW150914''},
                         arXiv: 1602.04460 [astro-ph.HE].

\bibitem{int} V. Savchenko {\it et al.},
             {\it ``INTEGRAL upper limits on gamma-ray emission associated with the gravitational wave event GW150914''},
             arXiv: 1602.04180 [astro-ph.HE].

\bibitem{des} DES Collaboration,
             {\it ``A Dark Energy Camera Search for an Optical Counterpart to the First Advanced LIGO Gravitational Wave Event GW150914''},
             arXiv: 1602.04198 [astro-ph.CO].

\bibitem{fermiLAT} Fermi-LAT Collaboration,
                  {\it ``Fermi-LAT Observations of the LIGO event GW150914''},
                  arXiv: 1602.04488 [astro-ph.HE].

\bibitem{swift}  Swift Collaboration,
  {\it ``Swift follow-up of the Gravitational Wave source GW150914''},
  arXiv: 1602.03868 [astro-ph.HE].
  %%CITATION = ARXIV:1602.03868;%%

\bibitem{agile}  M. Tavani {\it et al.},
  {\it ``AGILE Observations of the Gravitational Wave Event GW150914''},
  arXiv: 1604.00955 [astro-ph.HE].
  %%CITATION = ARXIV:1604.00955;%%

\bibitem{xmm} E. Troja, A.M. Read, A. Tiengo and R. Salvaterra,
  {\it ``XMM-Newton Slew Survey observations of the gravitational wave event GW150914''},
  arXiv: 1603.06585 [astro-ph.HE].
  %%CITATION = ARXIV:1603.06585;%%

\bibitem{rew1} B.P. Abbott {\it et al.} 
[for the LIGO Scientific and the Virgo and Pathfinder for the Australian Square Kilometer Array
and for the BOOTES and Survey for the Dark Energy and the Dark Energy Camera GW-EM s and for 
the Fermi GBM and for the Fermi LAT and TeAm for the GRAvitational Wave Inaf and for the INTEGRAL 
and Factory for the Intermediate Palomar Transient and Network for the InterPlanetary
and for the J-GEM and Survey for the La Silla--QUEST and for the Liverpool Telescope and 
Array for the Low Frequency and for the MASTER and
for the MAXI and Array for the Murchison Wide-field and for the Pan-STARRS and for the 
PESSTO and for the Pi of the Sky and for the
SkyMapper and for the Swift and TAROT for the and C2PU and for the TOROS and for the VISTA Collaborations],
 {\it ``Supplement: Localization and broadband follow-up of the gravitational-wave transient GW150914''},
  arXiv: 1604.07864 [astro-ph.HE].
  %%CITATION = ARXIV:1604.07864;%%

\bibitem{icecube} ANTARES and IceCube and LIGO Scientific and Virgo Collaborations,
                 {\it ``High-energy Neutrino follow-up search of Gravitational Wave Event GW150914 with ANTARES and IceCube''},
                 arXiv: 1602.05411 [astro-ph.HE].

\bibitem{GW151226} B.~P.~Abbott {\it et al.} [LIGO Scientific and Virgo Collaborations],
  {\it ``GW151226: Observation of Gravitational Waves from a 22-Solar-Mass Binary Black Hole Coalescence''},
  Phys.\ Rev.\ Lett.\  {\bf 116} (2016) no.24,  241103
  doi:10.1103/PhysRevLett.116.241103
  [arXiv:1606.04855 [gr-qc]].
  %%CITATION = doi:10.1103/PhysRevLett.116.241103;%%

\bibitem{Fermi_GW151226} J.~L.~Racusin {\it et al.} [Fermi-LAT Collaboration],
  {\it ``Searching the Gamma-ray Sky for Counterparts to Gravitational Wave Sources: 
  Fermi GBM and LAT Observations of LVT151012 and GW151226''}, 
  arXiv:1606.04901 [astro-ph.HE].
  %%CITATION = ARXIV:1606.04901;%%

\bibitem{Opticsi_1_GW151226} P.~S.~Cowperthwaite {\it et al.} [DES Collaboration],
  {\it ``A DECam Search for an Optical Counterpart to the LIGO Gravitational Wave Event GW151226''},
  [arXiv:1606.04538 [astro-ph.HE]].
  %%CITATION = ARXIV:1606.04538;%%

\bibitem{Opticsi_2_GW151226} S.~J.~Smartt {\it et al.},
  {\it ``A search for an optical counterpart to the gravitational wave event GW151226''},
  arXiv:1606.04795 [astro-ph.HE].
  %%CITATION = ARXIV:1606.04795;%%

\bibitem{ligoGR} LIGO Scientific and Virgo Collaborations,
                {\it ``Tests of general relativity with GW150914''},
                arXiv: 1602.03841 [gr-qc].

\bibitem{sibiryacovVelocity} D. Blas, M.M. Ivanov, I. Sawicki and S. Sibiryakov,
                            {\it ``On constraining the speed of gravitational waves following GW150914''},
                            arXiv: 1602.04188 [gr-qc].

\bibitem{shapiro}  E.O. Kahya and S. Desai,
  {\it ``Constraints on frequency-dependent violations of Shapiro delay from GW150914''},
  Phys. Lett. {\bf B 756} (2016) 265, arXiv: 1602.04779 [gr-qc].
  %%CITATION = doi:10.1016/j.physletb.2016.03.033;%%

\bibitem{ellisLigoFermi} J. Ellis, N.E. Mavromatos and D.V. Nanopoulos,
                        {\it ``Comments on Graviton Propagation in Light of GW150914''},
                        arXiv: 1602.04764 [gr-qc].

\bibitem{Silk}  K. Kotera and J. Silk, 
    {``Ultrahigh Energy Cosmic Rays and Black Hole Mergers''},
    arXiv: 1602.06961 [astro-ph.HE].
  %%CITATION = ARXIV:1602.06961;%%

\bibitem{jap} R. Yamazaki, K. Asano and Y. Ohira,
  {\it ``Electromagnetic Afterglows Associated with Gamma-Ray Emission Coincident with Binary Black Hole Merger Event GW150914''},
  arXiv: 1602.05050 [astro-ph.HE].
  %%CITATION = ARXIV:1602.05050;%%

\bibitem{plasma1} M. Milosavljevic and E.S. Phinney,
  {\it ``The Afterglow of massive black hole coalescence''},
  Astrophys. J. {\bf 622} (2005) L93, arXiv: astro-ph/0410343.
  %%CITATION = doi:10.1086/429618;%%


\bibitem{afterglow-1}
  S.M. O'Neill, M.C. Miller, T. Bogdanovic, C.S. Reynolds and J. Schnittman,
  {\it ``Reaction of Accretion Disks to Abrupt Mass Loss During Binary Black Hole Merger''},
  Astrophys. J. {\bf 700} (2009) 859, arXiv: 0812.4874 [astro-ph].
  %%CITATION = doi:10.1088/0004-637X/700/1/859;%%
  
\bibitem{afterglow-2}
  C. Palenzuela, M. Anderson, L. Lehner, S.L. Liebling and D. Neilsen,
  {\it ``Stirring, not shaking: binary black holes' effects on electromagnetic fields''},
  Phys. Rev. Lett. {\bf 103} (2009) 081101, arXiv: 0905.1121 [astro-ph.HE].
  %%CITATION = doi:10.1103/PhysRevLett.103.081101;%%

\bibitem{afterglow-3}
  C. Palenzuela, L. Lehner and S.L. Liebling,
  {\it ``Dual Jets from Binary Black Holes''},
  Science {\bf 329} (2010) 927, arXiv: 1005.1067 [astro-ph.HE].
  %%CITATION = doi:10.1126/science.1191766;%%
  
\bibitem{afterglow-4}
   P. Moesta, D. Alic, L. Rezzolla, O. Zanotti and C. Palenzuela,
  {\it ``On the detectability of dual jets from binary black holes''},
  Astrophys. J. {\bf 749} (2012) L32, arXiv: 1109.1177 [gr-qc].
  %%CITATION = doi:10.1088/2041-8205/749/2/L32;%%
  
\bibitem{afterglow-5}
  T. Bode, T. Bogdanovic, R. Haas, J. Healy, P. Laguna and D. Shoemaker,
  {\it ``Mergers of Supermassive Black Holes in Astrophysical Environments''},
  Astrophys. J. {\bf 744} (2012) 45, arXiv:1101.4684 [gr-qc].
  %%CITATION = doi:10.1088/0004-637X/744/1/45;%%

\bibitem{afterglow-6}
  B. Giacomazzo, J.G. Baker, M.C. Miller, C.S. Reynolds and J.R. van Meter,
  {\it ``General Relativistic Simulations of Magnetized Plasmas around Merging Supermassive Black Holes''},
  Astrophys. J. {\bf 752} (2012) L15, arXiv: 1203.6108 [astro-ph.HE].
  %%CITATION = doi:10.1088/2041-8205/752/1/L15;%%

\bibitem{afterglow-7}
  R. Gold, V. Paschalidis, Z.B. Etienne, S.L. Shapiro and H.P. Pfeiffer,
  {\it ``Accretion disks around binary black holes of unequal mass:
  General relativistic magnetohydrodynamic simulations near decoupling''},
  Phys. Rev. {\bf D 89} (2014) 064060, arXiv: 1312.0600 [astro-ph.HE].
  %%CITATION = doi:10.1103/PhysRevD.89.064060;%%

\bibitem{firsJets} R.D. Blandford and R.L. Znajek,
  {\it ``Electromagnetic extractions of energy from Kerr black holes''},
  Mon. Not. Roy. Astron. Soc.  {\bf 179} (1977) 433.
  %%CITATION = MNRAA,179,433;%%

\bibitem{lyut1} M. Lyutikov,
  {\it ``Electromagnetic power of merging and collapsing compact objects''},
  Phys. Rev. {\bf D 83} (2011) 124035, arXiv: 1104.1091 [astro-ph.HE].
  %%CITATION = doi:10.1103/PhysRevD.83.124035;%%

\bibitem{lyut2} M. Lyutikov,
  {\it ``Fermi GBM signal contemporaneous with GW150914 - an unlikely association''},
  arXiv: 1602.07352 [astro-ph.HE].
  %%CITATION = ARXIV:1602.07352;%%

\bibitem{masaros} K. Murase, K. Kashiyama, P. Meszaros, I. Shoemaker and N. Senno,
  {\it ``Ultrafast Outflows from Black Hole Mergers with a Mini-Disk''},
  Astrophys. J. {\bf 822} (2016) L9, arXiv: 1602.06938 [astro-ph.HE].
  %%CITATION = doi:10.3847/2041-8205/822/1/L9;%%


\bibitem{vachaspati1} T. Vachaspati,
  {\it ``Black Stars and Gamma Ray Bursts''},
  arXiv: 0706.1203 [astro-ph].
  %%CITATION = ARXIV:0706.1203;%%

\bibitem{blackStar1} S.L. Shapiro and S.A. Teukolsky,
  {\it ``Black holes, white dwarfs, and neutron stars: The physics of compact objects''}
  (Wiley, NY 1983).

\bibitem{blackStar2}  S.N. Zhang, Y. Liu, S. Yi, Z. Dai and C. Huang,
  {\it ``How to identify the gravitational wave events from merging stellar mass black binaries like GW150914?''},
  arXiv: 1604.02537 [gr-qc].
  %%CITATION = ARXIV:1604.02537;%%

\bibitem{genBZ} P.~Veres, R.~D.~Preece, A.~Goldstein, P.~Meszaros, E.~Burns and V.~Connaughton,
  {\it ``Gravitational wave observations may constrain gamma-ray burst models: the case of GW 150914 - GBM ''},
  arXiv:1607.02616 [astro-ph.HE].
  %%CITATION = ARXIV:1607.02616;%%

\bibitem{loeb} A. Loeb,
  {\it ``Electromagnetic Counterparts to Black Hole Mergers Detected by LIGO''},
  Astrophys. J. {\bf 819} (2016)  L21, arXiv: 1602.04735 [astro-ph.HE].
  %%CITATION = doi:10.3847/2041-8205/819/2/L21;%%

\bibitem{reconnection} F. Fraschetti,
  {\it ``Possible role of magnetic reconnection in the electromagnetic counterpart of binary black hole merger''},
  arXiv: 1603.01950 [astro-ph.HE].
  %%CITATION = ARXIV:1603.01950;%%

\bibitem{singularity}  D. Malafarina and P.S. Joshi,
  {\it ``Electromagnetic Counterparts to Gravitational Waves from Black Hole Mergers and Naked Singularities''},
  arXiv: 1603.02848 [gr-qc].
  %%CITATION = ARXIV:1603.02848;%%

\bibitem{merabRS} M. Gogberashvili,
                 {\it ``Hierarchy problem in the shell universe model''},
                 Int. J. Mod. Phys. {\bf D 11} (2002) 1635, arXiv: hep-ph/9812296.

\bibitem{RS} L. Randall and R. Sundrum,
            {\it ``A Large mass hierarchy from a small extra dimension''},
            Phys. Rev. Lett.  {\bf 83} (1999) 3370, arXiv: hep-ph/9905221;
            {\it ``An Alternative to compactification''},
            Phys. Rev. Lett. {\bf 83} (1999) 4690, arXiv: hep-th/9906064.

\bibitem{merabShellUniverse} M. Gogberashvili,
                            {\it ``Our world as an expanding shell''},
                            Europhys. Lett. {\bf 49} (2000) 396, arXiv: hep-ph/9812365.

\bibitem{Bubble} M. Gogberashvili,
                {\it ``Acceleration of a Spherical Brane-Universe''},
                Phys. Lett. {\bf B 636} (2006) 147, arXiv: gr-qc/0511039.

\bibitem{Go-Ma} M. Gogberashvili and M. Maziashvili,
               {\it ``Dark matter in the framework of shell-universe''},
               Gen. Rel. Grav. {\bf 37} (2005) 1129, arXiv: astro-ph/0404117.

\bibitem{Plank} Planck Collaboration,
               {\it ``Planck 2015 results. XIII. Cosmological parameters''}.
               arXiv: 1502.01589 [astro-ph.CO].

\end{thebibliography}
\end{document}